\documentclass[]{spie}  

\usepackage{amsmath,amsfonts,amssymb}
\usepackage{graphicx}
\usepackage{caption}
\usepackage{subcaption}
\usepackage[colorlinks=true, allcolors=blue]{hyperref}
\usepackage{float}

\title{Simons Observatory large aperture telescope receiver design overview}

\author[a]{Ningfeng Zhu}
\author[a]{John L. Orlowski-Scherer}
\author[a]{Zhilei Xu}
\author[b]{Aamir Ali}
\author[c]{Kam S. Arnold}
\author[b,d]{Peter C. Ashton}
\author[a]{Gabriele Coppi}
\author[a]{Mark J. Devlin}
\author[a]{Simon Dicker}
\author[c]{Nicholas Galitzki}
\author[e]{Patricio A. Gallardo}
\author[f]{Shawn W. Henderson}
\author[g]{Shuay-Pwu Patty Ho}
\author[h]{Johannes Hubmayr}
\author[c]{Brian Keating}
\author[b,d,i]{Adrian T. Lee}
\author[a]{Michele Limon}
\author[g]{Marius Lungu}
\author[j]{Philip D. Mauskopf}
\author[k]{Andrew J. May}
\author[l]{Jeff McMahon}
\author[e]{Michael D. Niemack}
\author[k]{Lucio Piccirillo}
\author[m]{Giuseppe Puglisi}
\author[d,n]{Mayuri Sathyanarayana Rao}
\author[o]{Maria Salatino}
\author[c]{Max Silva-Feaver}
\author[l]{Sara M. Simon}
\author[g]{Suzanne Staggs}
\author[a,p]{Robert Thornton}
\author[h]{Joel N. Ullom}
\author[e]{Eve M. Vavagiakis}
\author[b]{Benjamin Westbrook}
\author[q]{Edward J. Wollack}

\affil[a]{Department of Physics \& Astronomy, University of Pennsylvania, Philadelphia, PA, USA}

\affil[b]{Department of Physics, University of California, Berkeley, Berkeley, CA, USA}

\affil[c]{Department of Physics, University of California San Diego, La Jolla, CA, USA}

\affil[d]{Physics Division, Lawrence Berkeley National Laboratory, Berkeley, CA, USA}

\affil[e]{Department of Physics, Cornell University, Ithaca, NY, USA}

\affil[f]{Kavli Institute for Particle Astrophysics and Cosmology, SLAC National Accelerator Laboratory, Menlo Park, CA, USA}

\affil[g]{Department of Physics, Princeton University, Princeton, NJ, USA}

\affil[h]{Quantum Sensors Group, NIST, Boulder, CO, USA}

\affil[i]{Radio Astronomy Laboratory, University of California, Berkeley, Berkeley, CA, USA}

\affil[j]{School of Earth and Space Exploration and Department of Physics, Arizona State University, Tempe, AZ, USA}

\affil[k]{School of Physics and Astronomy, University of Manchester, Manchester, UK}

\affil[l]{Department of Physics, University of Michigan, Ann Arbor, Michigan, USA}

\affil[m]{Department of Physics, Stanford University, Stanford, CA, USA}

\affil[n]{Astronomy \& Astrophysics, Raman Research Institute, Bangalore, India}

\affil[o]{Astroparticle and Cosmology (APC) laboratory, Paris Diderot University, Paris, France}

\affil[p]{Department of Physics \& Engineering, West Chester University of Pennsylvania, West Chester, PA, USA}

\affil[q]{NASA/Goddard Space Flight Center, Greenbelt, MD, USA}

\authorinfo{Further author information: (Send correspondence to Ningfeng Zhu)\\N.Z: E-mail: nin@sas.upenn.edu}

\begin{document} 
\maketitle

\begin{abstract}

The Simons Observatory (SO) will make precision temperature and polarization measurements of the cosmic microwave background (CMB) using a series of telescopes which will cover angular scales between one arcminute and tens of degrees and sample frequencies between 27 and 270 GHz. Here we present the current design of the large aperture telescope receiver (LATR), a 2.4\,m diameter cryostat that will be mounted on the SO 6\,m telescope and will be the largest CMB receiver to date. The cryostat size was chosen to take advantage of the large focal plane area having high Strehl ratios, which is inherent to the Cross-Dragone telescope design. The LATR will be able to accommodate thirteen optics tubes, each having a 36\,cm diameter aperture and illuminating several thousand transition-edge sensor (TES) bolometers. This set of equipment will provide an opportunity to make measurements with unparalleled sensitivity. However, the size and complexity of the LATR also pose numerous technical challenges. In the following paper, we present the design of the LATR and include how we address these challenges. The solutions we develop in the process of designing the LATR will be informative for the general CMB community, and for future CMB experiments like CMB-S4.

\end{abstract}


\section{INTRODUCTION}
\label{sec:intro}  

The cosmic microwave background (CMB) has become one of the most powerful probes of the early universe. Measurements of temperature anisotropies on the level of ten parts per million have brought cosmology into a precision era, and have placed tight constraints on the fundamental properties of the universe\cite{Samtleben2007}. Beyond temperature anisotropies, CMB polarization anisotropies not only enrich our understanding of our cosmological model, but could potentially provide clues to the very beginning of the universe via the detection (or non-detection) of primordial gravitational waves.  A number of experiments have made and are continuing to refine measurements of the polarization anisotropy.  However, these experiments are typically dedicated to a relatively restricted range of angular scales, e.g., large angular scales (tens of degrees) or high resolution/small angular scales (on the order of one arcminute). To provide a complete picture of cosmology, both large and small angular scales are important. Ideally these measurements would be made from the same observing site so that the widest range of angular scales can be probed, at multiple frequencies, on the same regions of the sky. This is the goal of the Simons Observatory (SO).  

   \begin{figure}[H]
    	\begin{center}
        \begin{tabular}{c}
        \includegraphics[width = 0.8\linewidth]{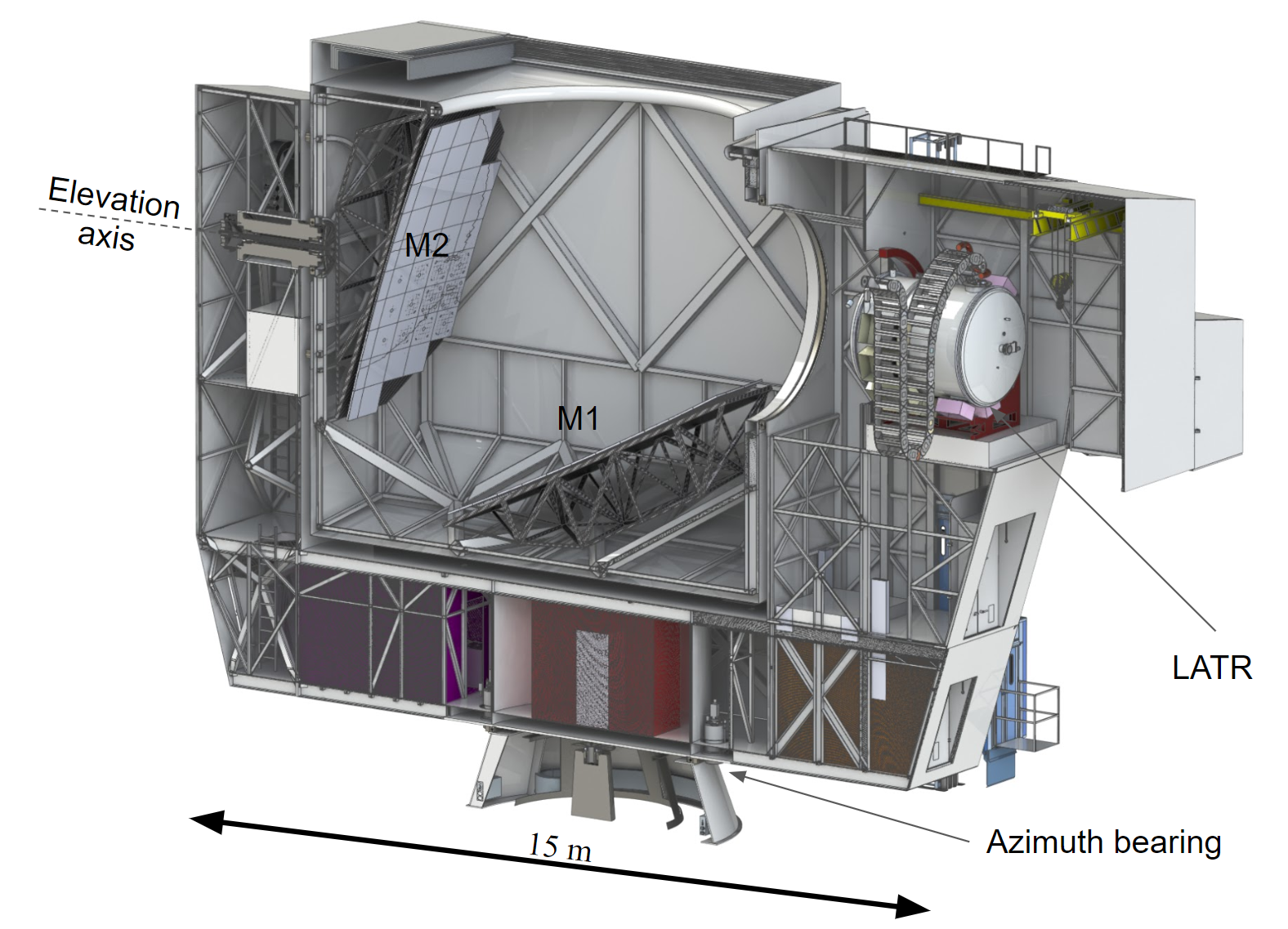}
        \end{tabular}
        \end{center}
    	\caption{A cross-section showing how the LATR couples to the SO 6\,m telescope.  In the orientation shown, light enters the telescope from the top.  It then reflects off the 6\,m primary (M1) and 6\,m secondary (M2) before being directed to the LATR receiver. The receiver always operates in the horizontal orientation shown, and is not coupled to the telescope elevation structure.  Therefore, a separate mechanism is used to rotate the receiver about its long axis as the telescope elevation structure moves M1 and M2 in rotation.} 
        \label{fig:telescope}
    \end{figure}

SO will field a 6\,m Cross-Dragone large aperture telescope (LAT) coupled to the large aperture telescope receiver (LATR, as shown in Fig.~\ref{fig:telescope}). The LAT is designed to have a large FOV\cite{Niemack2016,Parshley2018} capable of supporting a cryostat with up to 19 LATR-like optics tubes. To reduce the development risk of such large cryostat, the LATR is designed to accommodate only up to 13 optics tubes. During the initial deployment, we plan to deploy 7 optics tubes with 3 detector wafers in each for a total of approximately 35,000 detectors, primarily at 90/150\,GHz. It should be noted that each optics tube could be upgraded to support 4.5 wafers for a $\sim$50\% increase in the number of detectors per optics tube. With this upgrade and the deployment of 19 optics tubes, the LAT could support roughly 145,000 detectors at 90/150 GHz. SO will also deploy an array of half-meter small aperture telescopes (SATs) coupled to an additional 30,000 detectors\cite{Galitzki2018}. The unique combination of telescopes in a single CMB observatory, which will be located in Chile\textsc{\char13}s Atacama Desert at an altitude of 5190\,m, will allow us to sample a wide range of angular scales over a common survey area. 

In this paper, we present the current design of the LATR. In Sec.~\ref{sec:mech}, we will present the mechanical design of each temperature stage and the design philosophy behind each element. In Sec.~\ref{sec:cryo_design}, we will discuss a few key cryogenic components, including cryo-coolers, heat pipes, heat switches, thermometers, and infrared (IR) blocking filters\cite{Tucker2006,Ade2006}  that are being considered for the LATR. We will also cover simulation and testing plans for these components. Finally, in Sec.~\ref{sec:detector}, we will briefly discuss the detectors and readout strategy. The challenges we faced when designing the LATR are unique in that we are building the largest and most complex CMB receiver to date. The solutions to these challenges will provide a critical stepping stone for the next generation CMB experiments, in particular, CMB-S4\cite{Abazajian2016,Abitbol2017}.

\section{LATR Mechanical Design Overview}
\label{sec:mech}

    \begin{figure}[H]
    	\centering
        \includegraphics[width = 0.85\linewidth]{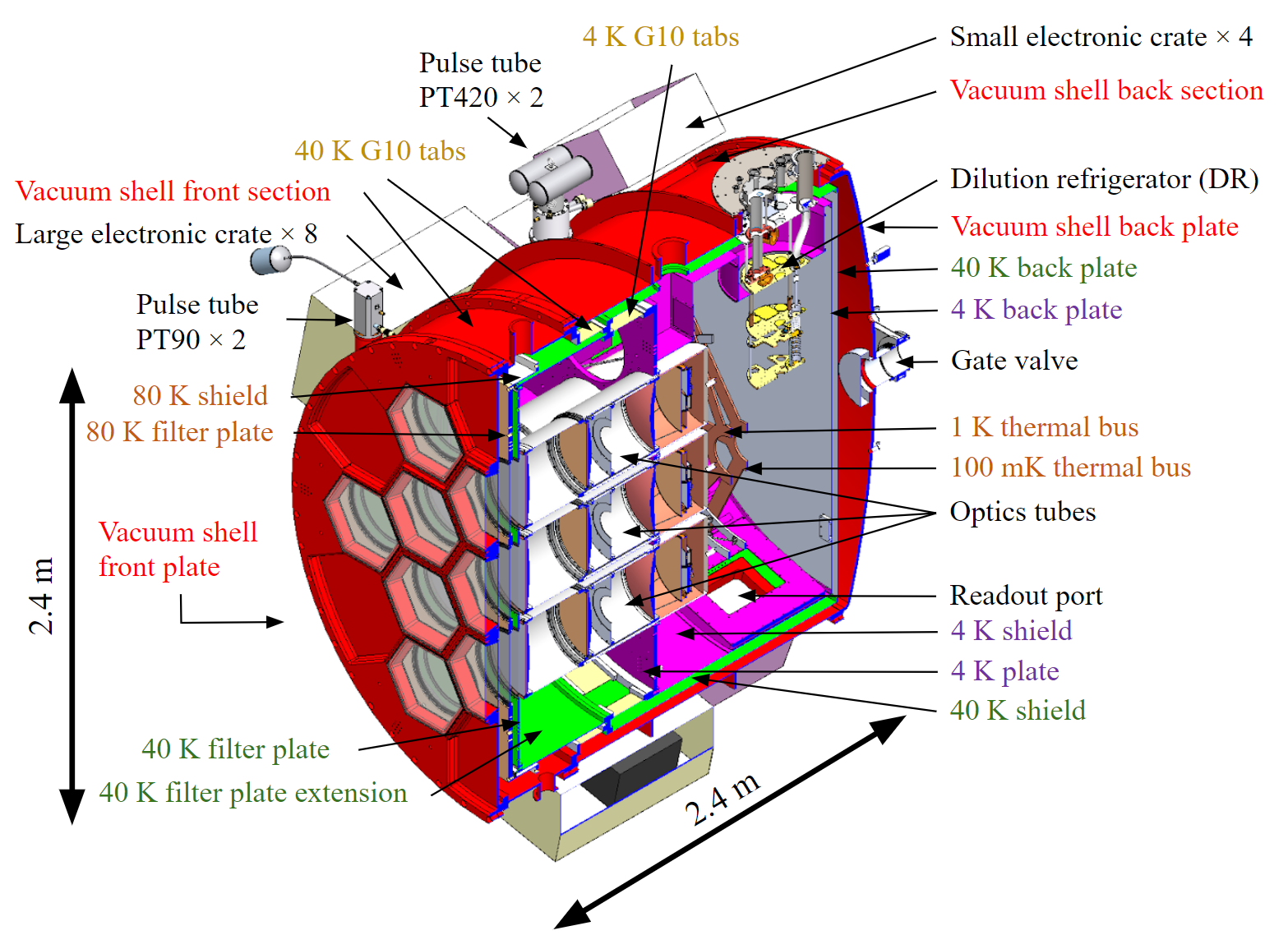}
    	\caption{Cross-section showing the internal structure of the LATR. For clarity, the optical elements are not labeled. However, in the orientation shown, the light enters from the left, goes through 300\,K window, 300\,K IR blocking filter, 80\,K IR blocking filter, 80\,K alumina filter, 40\,K IR blocking filter, and then enters the optics tube.}
        \label{fig:overview}
    \end{figure}

	The LATR is a 2.4\,m diameter cryostat designed to couple to the LAT. After light reflects off the primary and the secondary mirrors, it is directed to the cold optics in the LATR, which re-image the telescope focus onto 100\,mK focal plane arrays. A cross-sectional view of the cryostat is shown in Fig.~\ref{fig:overview}. Challenges faced in the design of this ambitious cryostat include, but are not limited to, structural integrity, vibrational modes, buckling effects, and thermal contraction. To address some of these challenges, we utilize finite element analysis (FEA) to model these effects (FEA details are discussed in a separate paper\cite{Orlowski2018}), and adjust the design based on the analysis. In addition, the cryostat is designed to be mounted on a co-rotator that enable +/- 45$^{\circ}$ bore-sight rotation following the telescope elevation change. The co-rotator mount restricts the maximum mass of the cryostat to 6,000\,kg, which in turn forces us to reduce the cryostat mass wherever possible.

	The LATR will include six temperature stages: 300\,K, 80\,K, 40\,K, 4\,K, 1\,K, and 100\,mK. In this section we will focus on the major components of the cryostat and explain their design philosophy. In addition, we will briefly discuss the optical elements (lenses and filters) mounted at each stage\cite{Dicker2018}. The thermal simulations of optical elements will be discussed in Sec.~\ref{sec:ther}.

\subsection{300\,K Stage (Vacuum Shell)}

	\begin{figure}[h]
		\centering
        \includegraphics[width = \linewidth]{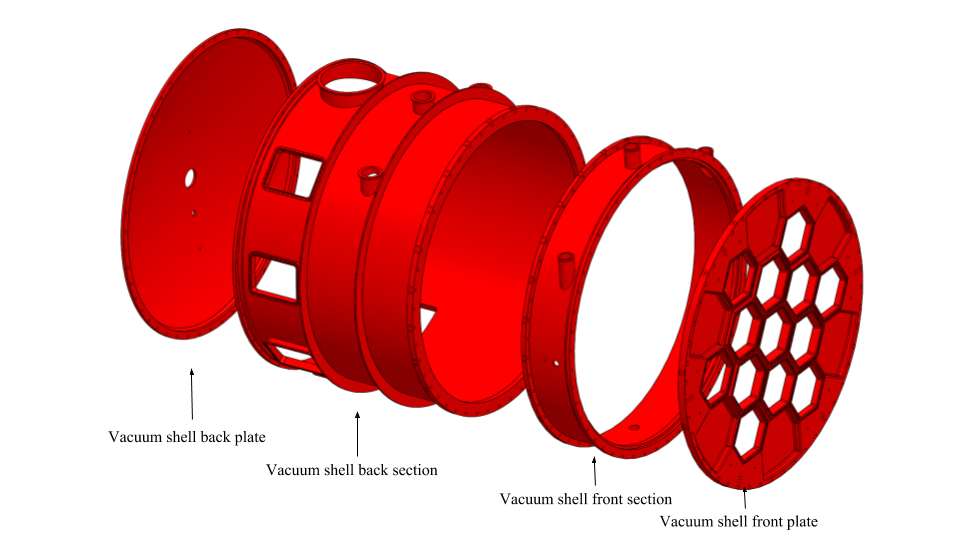}	
    	\caption{Exploded view showing the four major components of the vacuum shell. For scale, the assembled vacuum shell is 2.4\,m long and 2.4\,m in diameter. The front plate is reinforced where stress concentration is high, and weight-relieved where stress concentration is low. The front section shell is a 1.3\,cm thick cylinder and the back section shell is a 0.6\,cm thick cylinder. They are designed to be as weight efficient as possible to keep the cryostat within the total mass limit.}
    	\label{fig:vac_shell}
	\end{figure}

	The vacuum shell of the cryostat needs to withstand the atmospheric pressure since we need to reach a vacuum of at least $10^{-7}$ torr to minimize thermal conduction between stages. The model of the vacuum shell is shown in Fig.~\ref{fig:vac_shell}. The vacuum shell consists of four major components: the front plate, the front section shell, the back section shell, and the back plate. The front plate is a 6\,cm thick flat plate made of aluminum 6061 with thirteen densely packed hexagonal window cutouts to allow for maximum illumination on the detector arrays. The hexagonal window shape was chosen to match the focal plane module, which consists of three hexagonal-shaped arrays. In addition, the hexagonal window shape also reduces the structural stress on the front plate compared to an equivalent array of (circumscribed) circular cutouts. To ensure structural integrity of the shell, we conducted extensive FEA\cite{Orlowski2018} and had our results verified by an outside company, PVEng\footnote{Pressure Vessel Engineering Ltd, 120 Randall Dr b, Waterloo, ON N2V 1C6, Canada, https://pveng.com/}, to validate this aspect of the design. 

	We plan to use 0.3\,cm thick hexagonal windows made of anti-reflection coated ultra high molecular weight polyethylene. Each hexagonal window has its own vacuum-seal O-ring. This provides enough structural strength to withstand vacuum, as verified in a test chamber, and gives high in-band throughput. One double-sided IR blocking filter fabricated by Cardiff University is mounted on the back of the front plate behind each window to reduce the optical loading entering the cryostat.

\subsection{80\,K Stage}
    
	\begin{figure}[h]
		\centering
        \includegraphics[width = 0.6\linewidth]{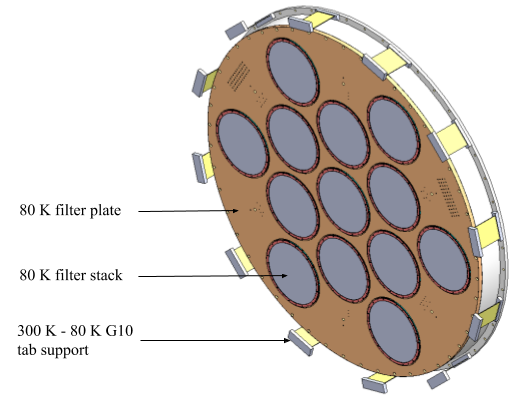}

    	\caption{Assembled model of the 80\,K stage. For scale the 80\,K plate is 2.3\,m in diameter. The combination of the IR blocking filter and alumina filter in the 80\,K filter stack will absorb most of the out-of-band optical power so that the 40\,K stage is not overwhelmed with optical loading. The 80\,K stage will absorb an estimated total of 100\,W from all loadings combined.}
    	\label{fig:80K_shell}
	\end{figure}
    
The 80\,K stage (as shown in Fig.~\ref{fig:80K_shell}) consists of a 2.3\,m diameter circular plate and a short 80\,K shield. A double-sided IR blocking filter and an anti-reflection coated alumina filter are mounted on the 80\,K stage. The double-sided IR blocking filter will reflect IR power and reduce the thermo-optical loading on the lower temperature stages; the alumina filter will act as an IR absorber as well as a prism to bend the off-axis beam back parallel to the long axis of the cryostat. The detailed design of this alumina filter is discussed in a separate paper\cite{Dicker2018}. 

The 80\,K stage is supported by a series of G10 tabs that are 14\,cm wide, 20\,cm long, and 0.2\,cm thick. This approach will accommodate the high differential thermal contraction (on the order of 1\,cm in diameter) between the 80\,K and the 300\,K stages during cooling. The G10 tabs provide some flexibility to bend radially inward during the cool-down process, while still holding the cold components in place. Extensive FEA has been performed to simulate the structural strength of the G10 tabs\cite{Orlowski2018}. The structural support of the 80\,K stage is located on the vacuum shell front section. Thus, the entire 80\,K stage can be installed or disassembled independently from the rest of the cryostat.

\subsection{40\,K Stage}

       \begin{figure}[H]
		\centering
        \includegraphics[width = 0.7\linewidth]{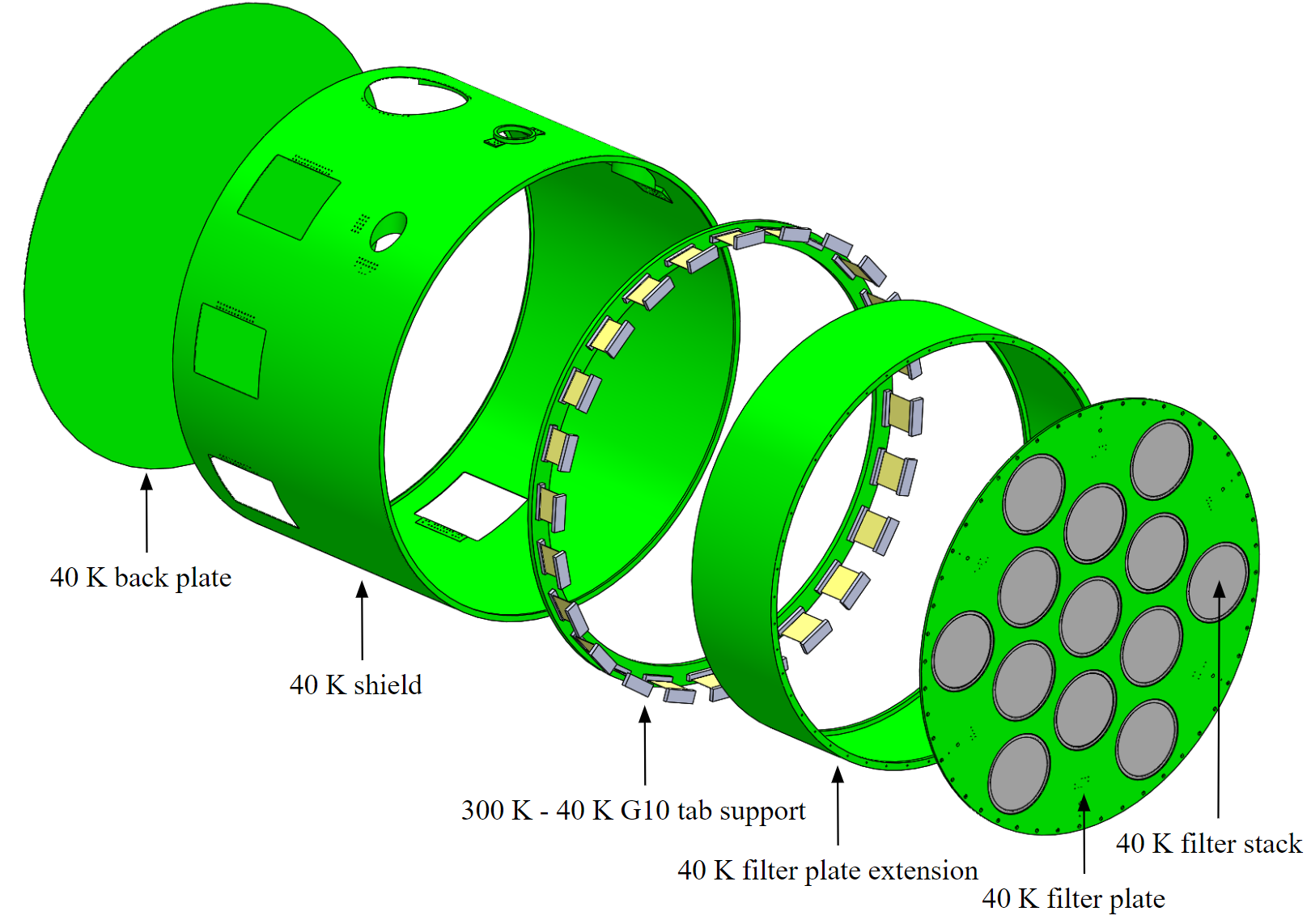}
    	\caption{The exploded view showing the 40\,K stage assembly. For scale, the 40\,K stage assembly is 2.1\,m in diameter and 2.0\,m long. As shown, the light enters from the right and goes through the IR blocking filters mounted on the 40\,K filter plate. The entirety of the 40\,K stage will be covered by multi-layer insulation (not shown in the model).}
    	\label{fig:40K_shell}
	\end{figure}
    
    	The 40\,K stage consists of a 2.1\,m diameter circular plate, a 40\,K radiation shield, a 40\,K filter extension tube, and a thin 40\,K radiation back plate, as shown in Fig.~\ref{fig:40K_shell}. Rather than being suspended from the 80\,K stage, it is instead suspended directly from the 300\,K vacuum shell. This approach greatly reduces the structural stress on the 80\,K stage G10 tabs, simplifies the assembly process, and adds little conductive load to the 40\,K stage compared to the cooling capacity available\cite{Orlowski2018}. Another double-sided IR blocking filter will be mounted on the 40\,K filter plate to further reduce the IR power to milliwatt level on lower temperature stages (per optics tube). 

    The structural support of the 40\,K stage is located on the vacuum shell back section. This provides a natural break in the vacuum shell for ease of assembly. There is a thick structural flange on either side of this break that minimizes stress on the vacuum shell itself. Similar to the 80\,K stage support, the 40\,K stage is supported by G10 tabs. Again, we did a thorough FEA validation to verify the structural strength of the G10 tab support\cite{Orlowski2018}. To reduce the radiative loading on the 40\,K stage, we plan to cover the entirety of the 40\,K assembly with multi-layer insulation (MLI), approximately 30 layers of aluminized mylar. To vent the inside of the 40\,K cavity, evacuation blocks were designed to allow air to escape during evacuation. This also prevents the fragile filters from bursting due to differential pressure.
    
\subsection{4\,K Stage}

    \begin{figure}[h]
		\centering
        \includegraphics[width = 0.7\linewidth]{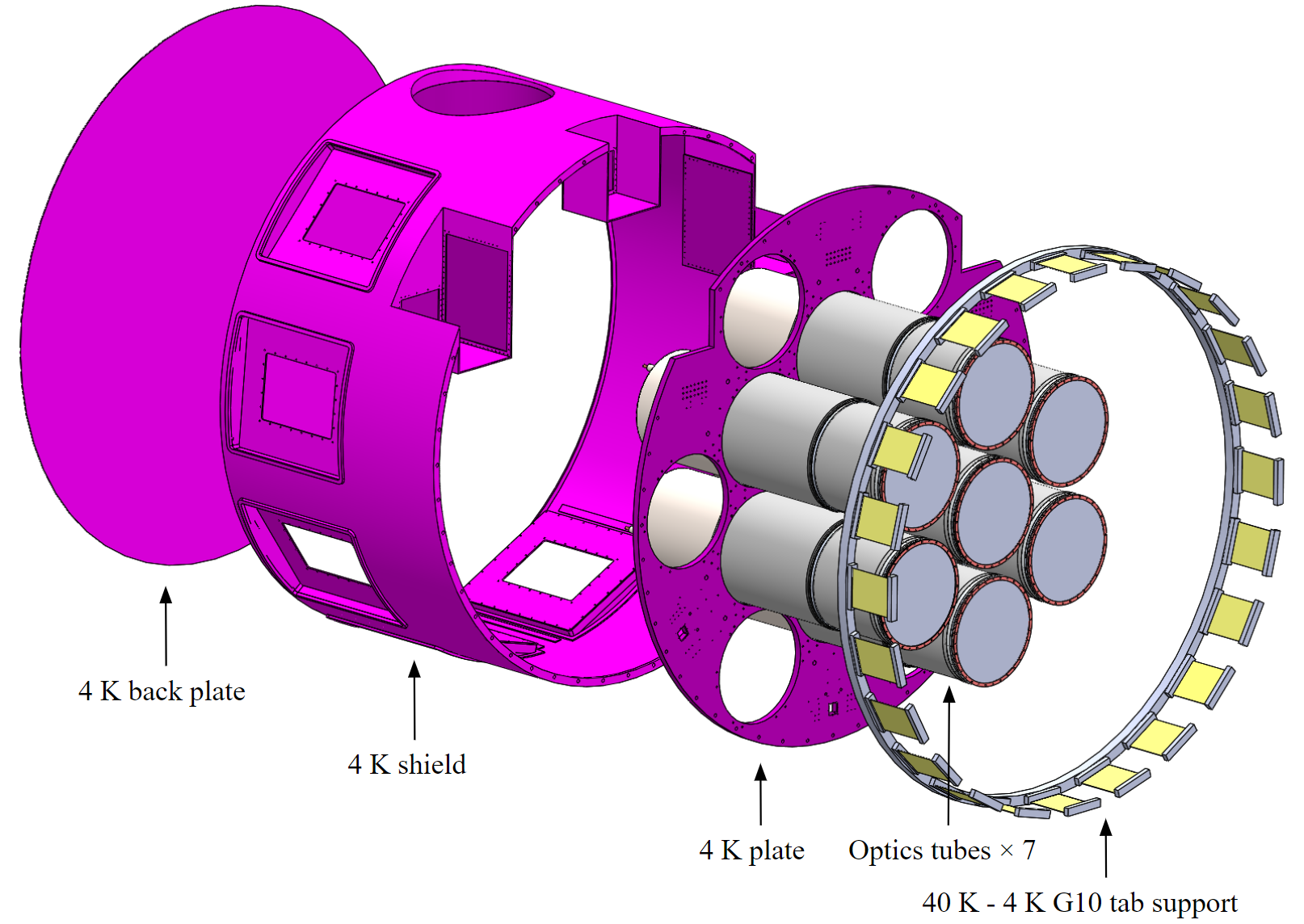}
   		\caption{An exploded view showing the 4\,K stage assembly. For scale the 4\,K plate is 2.1\,m in diameter. The optics tubes are mounted on the 4\,K plate and thus thermally coupled to the 4\,K stage. The radiative load from 40\,K to 4\,K is negligible, so no MLI is planned.}
    	\label{fig:4K_shell}
	\end{figure}

	The 4\,K stage consists of a 2.1\,m diameter, 2.5\,cm thick circular plate, a 4\,K radiation shield and a thin 4\,K back plate, as shown in Fig.~\ref{fig:4K_shell}. Similar to the 80\,K and 40\,K stages, the 4\,K stage assembly is also supported by G10 tabs. The main 4\,K plate is significantly thicker than the plates for the 40\,K or 80\,K stages since this structure supports the optics tubes (Sec.~\ref{sec:opt}), the most vital components of the cryostat. The 4\,K radiation shield is designed to have recesses for pulse tube refrigerator attachments without penetrating the 4\,K cavity, allowing for significant thermal strapping area while minimizing the chance of light leaks into the 4\,K cavity, where the 1\,K and the 100\,mK components reside. 

\subsection{Optics tube}
\label{sec:opt}

	 \begin{figure}[H]
		\centering
         \includegraphics[width = \linewidth]{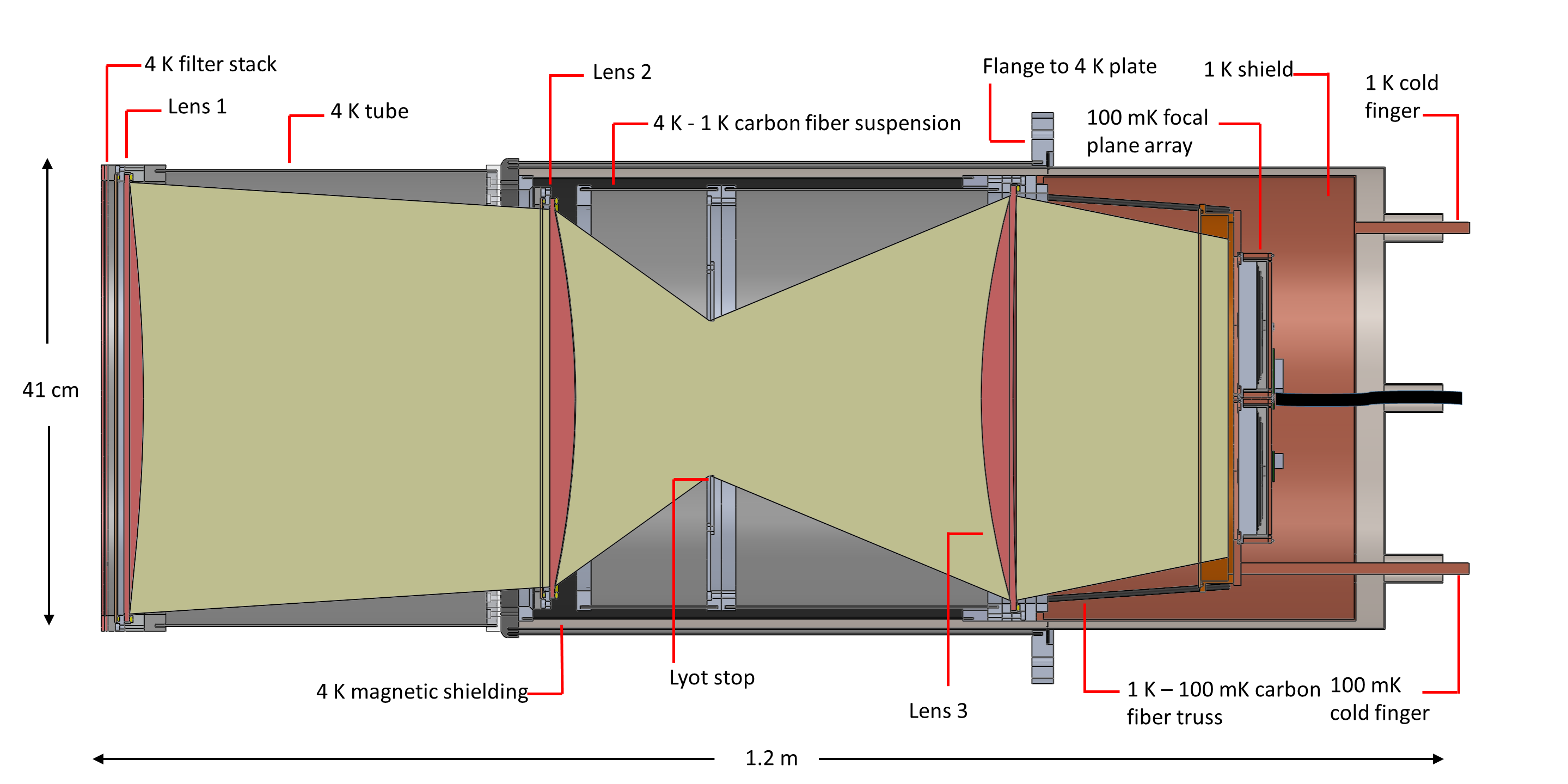}
         \caption{Cross section showing the internal structure of the optics tube and the projected light path between the 4\,K lens and the focal plane array.}
            \label{fig:optics_tube}
        \end{figure}

The optics tubes contain all optical and detector components between 4\,K and 100\,mK.  Each optics tube, a schematic of which is shown in Fig.~\ref{fig:optics_tube}, is designed to be self contained such that it can be installed as a single unit into the cryostat. An optics tube contains three silicon lenses that re-image the telescope focal plane onto an array of three hexagonal wafers (details are discussed in Dicker et al.\cite{Dicker2018}). The lenses feature a metamaterial anti-reflection (AR) coating similar to that described in Datta et al\cite{Datta13}. The first lens is at 4\,K, and is supported from the cryostat 4\,K cold plate by an approximately 1\,m long aluminum tube.  This 4\,K tube is designed to be thin, due to mass budget, but will be fabricated from either 1100-H14 or 6063-O aluminum as a compromise between strength and thermal conductivity.  A tube of magnetic shielding (A-4K)\footnote{A trademark of Vacuumschmelze GmbH in Hanau, Germany. Local Distributor: Amuneal Manufacturing Corporation, 4737 Darrah St., Philadelphia, PA 19124, USA, info@amuneal.com, (800)-755-9843.} will line the inside wall of the 4\,K tube.  The second lens, Lyot stop, and third lens are at 1\,K, and are supported from the 4\,K tube by a thermally-isolating carbon fiber tube.  The inside surfaces of all walls the beam travels along are blackened and/or include baffles (not shown).  The 100\,mK array assembly, whose details are discussed in Ho et al.\cite{Ho2018}, is suspended from the 1\,K structure by a carbon fiber truss. A second layer of magnetic shielding (not shown) surrounds the focal plane array.  Thermometry, cold fingers, and detector readout cables are fed through the back half of the A-4K magnetic shielding.

\subsection{1K and 100mK stages}

    	

	 \begin{figure}[H]
		\centering
         \includegraphics[width = 0.85\linewidth]{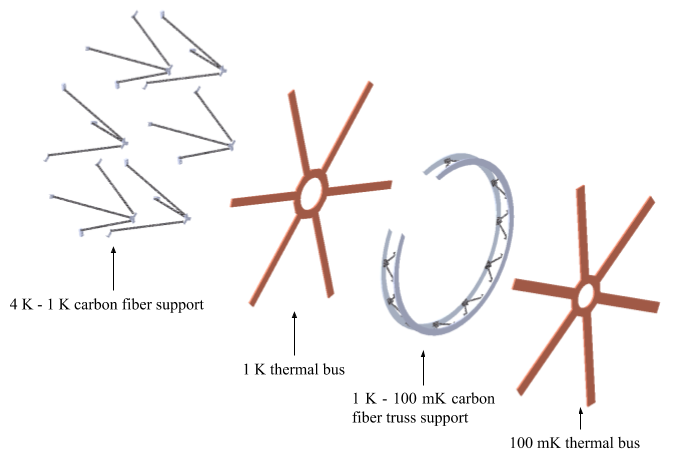}
         \caption{Exploded view showing the thermal bus assembly. The thermal buses are also shown in Fig.~\ref{fig:overview}. For scale the 1\,K - 100\,mK truss support is 1.0\,m in diameter. The thermal bus will be made of oxygen-free high-conductivity (OFHC) copper.}
            \label{fig:thermal_bus}
     \end{figure}

	To maximize the sensitivity of the TES bolometers, all detector components and readout chips will be cooled to 100\,mK. The cooling mechanism will be described in Sec.~\ref{sec:cryo}. The thermal conductivity of aluminum alloys is very low at 1\,K and below; thus, we use two snowflake structures made of OFHC copper to reduce thermal gradients on the 1\,K and 100\,mK stages. The solid model is shown in Fig.~\ref{fig:thermal_bus}. We performed  numerous thermal FEA and validated that the temperature gradient across either thermal bus is at the $\sim$10\,mK level, which satisfies our loading requirements\cite{Orlowski2018}. From the thermal bus, individual copper straps will extend down to attach to the 1\,K and 100\,mK structures inside each optics tube.

\section{LATR Cryogenics design OVERVIEW}
\label{sec:cryo_design}

\subsection{Cryogenic coolers}
\label{sec:cryo}
	The LATR temperature stages were chosen to reduce loading on subsequent stages to a level that base temperatures could be achieved with available cooling technology. A major concern when designing this system is that the cryostat could take more than a month to cool down to operation temperature due to its large thermal mass. The cryostat cool down calculations are discussed in Coppi et al\cite{Coppi2018}.  
    
	 The thermal load on the 80\,K stage is nearly 100\,W, dominated by IR radiation entering the thirteen windows. Thus, we decided to utilize two pulse tube PT90 coolers manufactured by Cryomech\footnote{Cryomech Inc, 113 Falso Dr, Syracuse, NY 13211, https://www.cryomech.com}, mounted on the vacuum shell front section near the 80\,K components (as shown in Fig.\ref{fig:overview}). To conduct heat from the PT90s to the 80\,K stage components, we will use copper braids as heat straps, which have been validated by thermal analyses\cite{Orlowski2018}. In the cryostat design, we left a third port for a potential third PT90 cooler in case the cool down time is significantly longer than calculated, or if the loading on the 80\,K stage is much higher than expected.
     
     The 40\,K and the 4\,K stages are cooled by two Cryomech PT420s, each of which provides 55\,W of cooling at 40\,K, and 2\,W of cooling at 4\,K. Similar to the 80\,K stage, we will use copper braids as heat straps. We also designed a third port to allow for a potential additional PT420 cooler due to similar concerns as at the 80\,K stage. 
     
     To maximize the sensitivity of our TES detectors, we use a dilution refrigerator manufactured by Bluefors\footnote{BlueFors Cryogenics, Arinatie 10, 00370 Helsinki, Finland, https://www.bluefors.com} to provide cooling at 1\,K and 100\,mK. The estimated cooling power is 400\,$\mu$W at 100\,mK. The dilution refrigerator is backed by its own PT420 cooler (not shown in Fig.~\ref{fig:overview}), which we will also use to help cool the cryostat 40\,K components.

\subsection{Nitrogen Heat pipe}

As discussed in Sec.~\ref{sec:mech}, a large amount of mass (approximately 1500\,kg) needs to be cooled to 4\,K or lower, nominally by the second stage (4\,K stage) of the PT420 pulse tube coolers, which naturally have less cooling power than the first stage (40\,K stage), especially at temperatures $>\,$100\,K. According to our thermal models, during the cryostat cooldown, when the 40\,K stage of a PT420 is near its base temperature, the 4\,K stage will still be at about 200\,K. Therefore, we would like to transfer heat efficiently between the two stages at high temperature and automatically shut this transfer off at lower temperatures ($\lesssim$ 40\,K). After consulting several cryogenic companies, we decided to develop a nitrogen heat pipe to address this issue.

A heat pipe using high pressure nitrogen gas~\cite{Shukla2015} can be used to facilitate thermal transport in a cryogenic system. In the LATR, the nitrogen heat pipe will be installed vertically, with the top end connected to the 40\,K stage of the PT420 and the bottom end connected to the 4\,K stage.  Once the top end is sufficiently cold, gaseous nitrogen in that part of the pipe liquefies and falls to the bottom, warmer end. When it then vaporizes at the bottom, it rises back up to recondense at the cold end, creating a circulation cycle. This circulation efficiently transfers heat from the bottom (initially warmer) end to the top (initially colder) end via the latent heat of nitrogen. As the cooling process continues, the temperature at both ends of the pipe continues to fall, and more nitrogen gas becomes liquefied. Eventually, very few nitrogen molecules are in the gaseous state. At this point, the circulation is effectively shut off, leaving only a negligible amount of conduction between the two stages (mostly through the stainless steel pipe). During cryostat cool-down process, the 4\,K stage effectively ``borrows" the cooling power from the 40\,K stage via the nitrogen heat pipe. Once the 4\,K stage is cooled to low enough temperature ($\lesssim$ 40\,K), the two ends of the pipe become mostly isolated.  The pulse tube second stage now has a comparatively easier time cooling the relevant cryostat components from $\sim$~40\,K to 4\,K.

\begin{figure}[H]
	\centering
	\includegraphics[width = \linewidth]{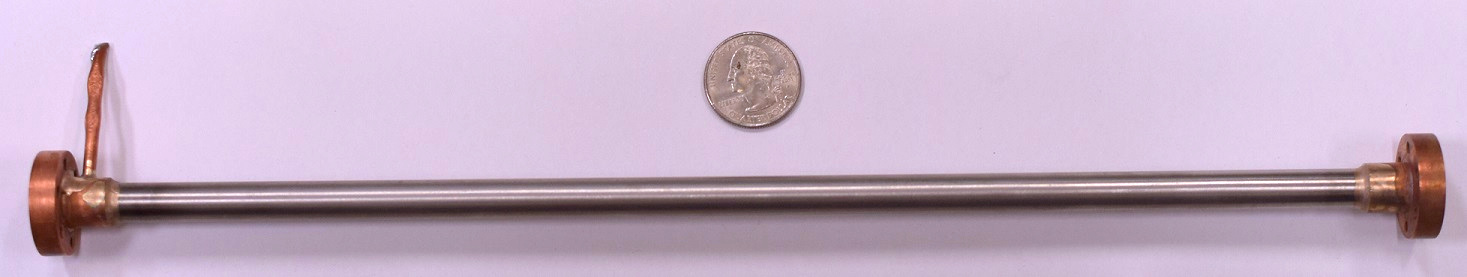}
    \caption{Nitrogen Heat Pipe. The middle section is a thin-walled stainless steel pipe. The two ends are caps made of OFHC copper. One of the caps has a capillary pipe for nitrogen gas charging. All of the individual pieces are hard soldered together.}
    \label{fig:heat_pipe}
\end{figure}

A photo of the nitrogen heat pipe we have fabricated is shown in Fig.~\ref{fig:heat_pipe}. Our heat pipes are charged to 1000~psi at room temperature, a pressure selected to have the condensing process triggered at $\sim$120~K. The heat pipes are made from stainless steel pipes with wall thickness of 0.3\,mm and outer diameter of 9.5\,mm. The mechanical design has been tested to hold the 1000\,psi internal pressure with a factor of safety of at least three. The thermal properties (such as on-state and off-state thermal conductivity) are currently being tested.

\subsection{Heat switch}

One strategy to reduce the cooldown time between the 4\,K, 1\,K, and 100\,mK stages is the implementation of heat switches. By temporarily creating a strong thermal link between the 1\,K stage and the 4\,K stage and between the 100\,mK stage and the 1\,K stage, the lower temperature stages may be pre-cooled and subsequently decoupled for further cooling to their base temperatures. In order to accomplish this, the heat switches must have high switching ratios (i.e., the conduction ratio between the ``open" and ``closed" states). Several different types of heat switches are currently under consideration and testing, including gas-gap and mechanical switches.

\begin{figure}[H]
\centering
\begin{minipage}{.45\textwidth}
  \centering
  \includegraphics[width=\linewidth]{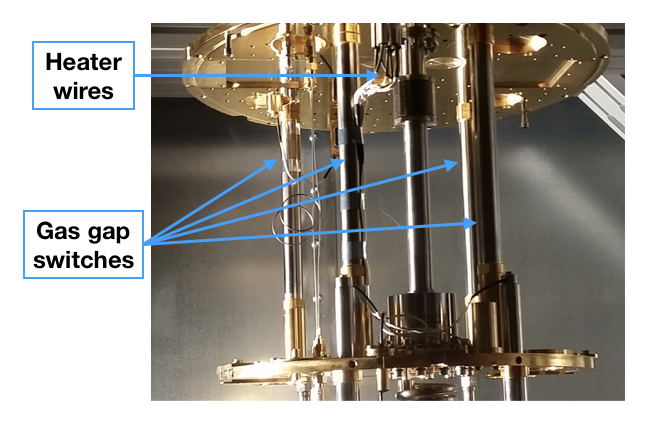}
  \caption{Gas gap heat switches shown mounted in a Bluefors LD250 cryostat; note the heater wires used to operate the absorption pump.}
  \label{fig:gasgapHS}
\end{minipage}%
\qquad
\begin{minipage}{.45\textwidth}
  \centering
  \includegraphics[width=\linewidth]{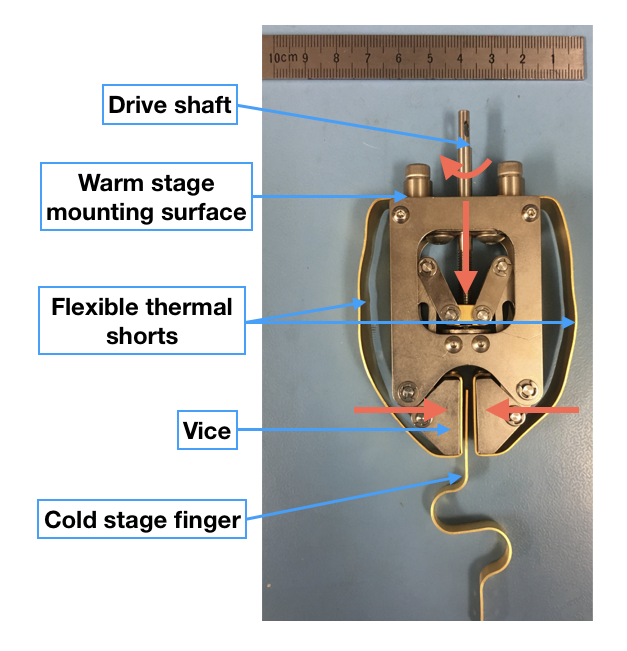}
  \caption{The HPD mechanical heat switch. The control box and the motor are not shown.}
  \label{fig:mechHS}
\end{minipage}
\end{figure}

Gas-gap switches operate through the use of a small helium adsorption pump which is able to actively fill or evacuate a small cavity between two copper rods inside a thin-walled stainless steel tube. The open conductance is reduced to the minimal conduction through the small cross-sectional area of the stainless steel tube. The closed conductance is several orders of magnitude larger, since the small gap between the cold and warm copper rods is filled with helium gas, which conducts heat across the gap efficiently. Several gas-gap switches of the type under consideration are shown in Fig.~\ref{fig:gasgapHS}. The switches, manufactured by Bluefors, are shown mounted in an LD250 dilution refrigerator cryostat.

Mechanical heat switches are operated by using some actuators to make or break a physical connection. In this way, zero open conductance is achieved, and very high closed conductance can be achieved as long as the required pressure can be developed at the interface. As noted elsewhere \cite{Coppi2018}, thermal boundary resistance is an increasingly significant issue at lower temperatures and, as such, extensive testing of several mechanical switches is currently underway. A mechanical switch of the type currently under consideration, manufactured by HPD\footnote{High Precision Devices, 4601 Nautilus Ct S, Boulder, CO 80301, USA, http://www.hpd-online.com}, is shown in Fig.~\ref{fig:mechHS}. This type of switch is closed by rotating the drive shaft to develop a high contact pressure between the vice and the cold stage finger; this couples the finger through flexible thermal shorts to the warm stage. 

\subsection{Thermometry}
	We plan to use 128 thermometers distributed among the various temperature stages to monitor the cryostat temperature during cool down and operation. The thermometers are strategically placed so that they can be used to read out temperature gradients on stages during integration and testing, as well as monitor temperature changes over time during nominal operation.
    
    We will have a combination of Ruthenium oxide sensors (ROXs) and diodes, depending on the temperature stage. Diodes will be used to read out temperatures at 4\,K and above, while ROXs will be used to read out temperatures at 1\,K and below. The distribution is included in Tab.~\ref{tab:thermometer}. Both types of sensors will be potted into copper bobbins for ease of installation and to protect the sensitive components. We plan to connect the thermometers using Omnetic\footnote{Omnetics Connector Corporation, 7260 Commerce Cir E, Minneapolis, MN 55432, USA, http://www.omnetics.com/} pin connectors. Each bobbin will have a two pin connector mounted on top, followed by a 2-wire to 4-wire transition cable that allows for 4-lead measurements on all thermometers.
	
    \begin{table}[h]
	\centering

	\begin{tabular}{|l|l|l|}
	\hline
	Stages & Diodes & ROXs \\ \hline
	80\,K    & 8      & 0    \\ \hline
	40\,K    & 16     & 0    \\ \hline
	4\,K     & 40     & 0    \\ \hline
	1\,K     & 0      & 32   \\ \hline
	100\,mK  & 0      & 32   \\ \hline
	Total  & 64     & 64   \\ \hline
	\end{tabular}
    \caption{Thermometer distribution across all temperature stages.}
	\label{tab:thermometer}
	\end{table}

\subsection{Thermal Optical Simulation }
\label{sec:ther}

\begin{figure}[H]
    	\centering
        \includegraphics[width = 0.6\linewidth]{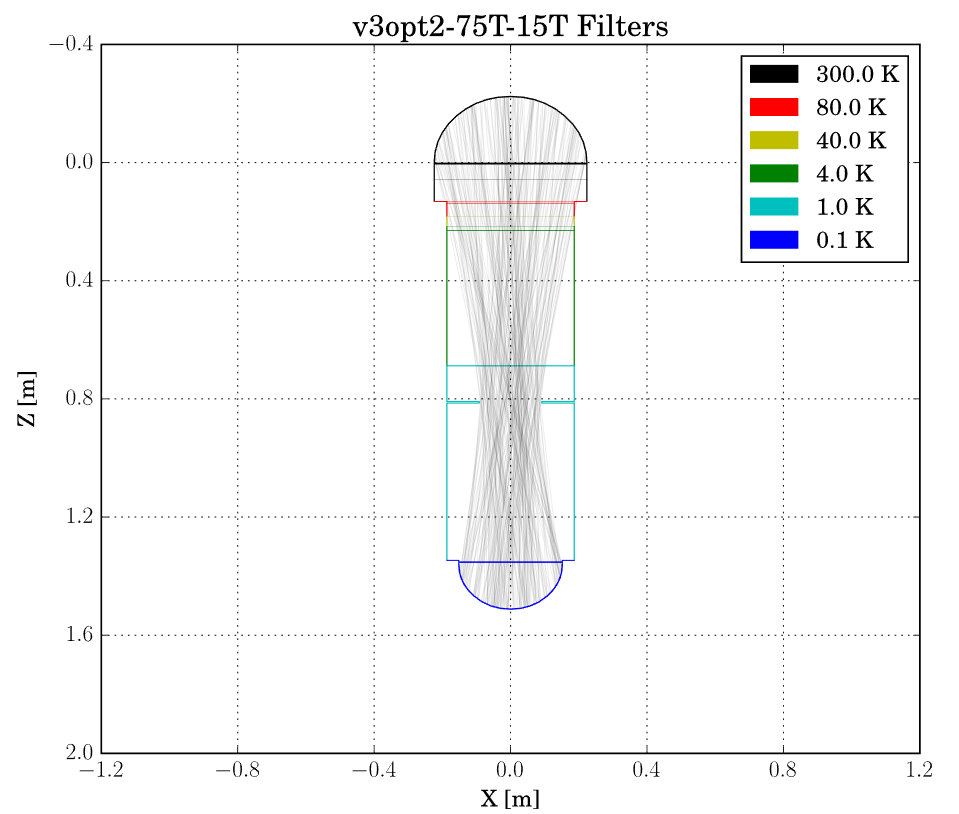}
    	\caption{Geometry of the LATR optics tube showing all filter elements and walls at each stage. The lines shown in the graph do not represent the actual ray trace; they merely highlight the location of the Lyot stop.}
        \label{fig:ray}
    \end{figure}

To calculate the thermal loading due to the optical elements inside the LATR, we created a custom Python package to model the thermal performance of all filter elements in a predefined radiative environment. The code estimates the total power emitted and absorbed at each temperature stage by performing a radiative transfer simulation using numerical ray optics. The circularly symmetrized geometry (shown in Fig.~\ref{fig:ray} for our current design) and the spectral properties of the optics tube walls and filters are used as inputs to this model. An additional output of this simulation is a set of radial temperature profiles for all filter elements (computed using temperature-dependent thermal conductivities of the filter materials). The resultant thermal loads for each stage are shown in Tab.~\ref{tab:load}, while the optimized filter stack and corresponding center temperatures are shown in Tab.~\ref{tab:filter_temp}.

	\begin{table}[h]
		\centering
			\begin{tabular}{|l|l|l|l|l|l|l|}
			\hline
			Temperature stage & 300\,K    & 80\,K    & 40\,K     & 4\,K      & 1\,K      & 100\,mK  \\ \hline
			Optical load (per optics tube)      & -3.93\,W & 3.90\,W & 2.0\,mW & 27.6\,mW & 29\,$\mu$W & 312\,nW \\ \hline
			\end{tabular}
		
		\caption{Table of thermal loading from the optical chain at each temperature stage. The negative loading at 300\,K represents the total power that enters the window.}
        \label{tab:load}
	\end{table}

	\begin{table}[h]
		\centering
			\begin{tabular}{|l|l|l|l|l|l|l|l|l|}
			\hline
			Temperature stage & 300\,K & 80\,K & 80\,K & 40\,K & 4\,K     & 4\,K & 1\,K & 100\,mK  \\ \hline
            Filter & IRB   & IRB & AF & IRB & IRB & LPE & LPE& LPE\\ \hline
			Filter center temperature      & 297\,K & 252\,K & 81\,K & 74\,K & 62\,K & 12\,K & 1.1\,K & 105\,mK \\ \hline
			\end{tabular}
		
		\caption{Table of optimized filter stack and center temperatures. ``IRB" stands for IR blocking filter, ``AF" stands for alumina filter, and ``LPE" stands for low pass edge filter.}
        \label{tab:filter_temp}
	\end{table}

The loading due to the optics is then incorporated into the master thermal spreadsheet, which combines all sources of thermal loads (conductive and radiative) in order to determine the total thermal power on each stage. The thermal spreadsheet will be discussed in a separate paper\cite{Orlowski2018}.

\section{Detectors and Readout}
\label{sec:detector}

The LATR will house approximately 35,000 TES bolometers coupled to either feedhorns or lenslets during initial deployment with seven of the planned thirteen optics tubes. To maximize the sensitivity of the detector arrays and achieve photon-limited noise over all optical frequency bands, we cool down the array modules to 100\,mK. We chose to distribute the frequency range of dichroic TES bolometers into three categories: LF (27\,GHz \& 39\,GHz), MF (93\,GHz \& 145\,GHz), and UHF (225\,GHz \& 280\,GHz), with each optics tube housing one dichroic pair for ease of filter and lens optimization. The current plan for initial development is to populate the inner seven optics tubes with one LF tube, four MF tubes, and two UHF tubes to maximize our science goals. These seven tubes will cover a sky fraction of $f_{sky}$ = 0.4. The array modules holding the lenslets/feedhorns, detector wafers, and cryogenic readout components will be discussed in a separate paper\cite{Ho2018}.

TES bolometer arrays will be read out using the Stanford Linear Accelerator Center (SLAC) superconducting microresonator radio frequency (SMuRF) electronics\cite{Kernasovskiy2018}. The traditional way of reading out TES bolometers is to utilize superconducting quantum interference devices (SQUIDs) with either time-division or frequency-domain multiplexing. However, the multiplexing factors of the current generation readout systems would require a significant amount of separate warm and cold readout chains and wires from 300\,K to 100\,mK, which adds nontrivial complexity to the cryogenic design. Thus the warm SMuRF electronics will be used to read out the LATR arrays using cold microwave SQUID multiplexers ($\mu$mux) manufactured by National Institute of Standards and Technology (NIST), which target much higher multiplexing factors ($\sim$2000 detectors per transmission line pair) \cite{Dober2017}.

The readout electronics for the LATR $\mu$mux system include four sets of readout harness inserts that will penetrate the vacuum shell (square cutouts in Fig.~\ref{fig:vac_shell}). These readout harnesses have components mounted on either the 40\,K shell or the 4\,K shell, including the low frequency wiring components, the radio frequency (RF) wiring, and RF hardware like amplifiers, attenuators, and DC blocks.  Mounted on the outside of the cryostat are twelve readout crates containing SMuRF electronics (as shown in Fig.~\ref{fig:overview}). These data acquisition crates are mounted directly to the vacuum vessel to minimize noise and electronic interference by maximizing their proximity to the detectors. They are also co-moving with the receiver to minimize movement in the low frequency and RF wiring from the readout crates to the inserts. Note that the penetrations of the readout harness inserts presented a challenge in that they significantly weakened the structural integrity of the vacuum shell, and therefore had to be carefully accounted for in the structural FEA of the vacuum shell\cite{Orlowski2018}. The large volume occupied by these crates also posed a challenge regarding space allocation given the large number of pulse tubes, the dilution refrigerator, and the telescope cable wrap.

\section{Summary and Status}
We have completed the major component design of the 2.4\,m receiver that will operate on the 6\,m millimeter-wave Simons Observatory LAT. The LATR will contain approximately 35,000 detectors (in the first seven of the planned thirteen optics tubes) operating at 100\,mK and sensitive to six frequency bands between 27 and 270\,GHz. With this set of equipment, the LATR will measure fundamental cosmological parameters of our universe with unparalleled sensitivity, find high redshift clusters through the Sunyaev-Zel'dovich effect, constrain properties of neutrinos, and seek signatures of dark matter through gravitational lensing. Manufacture of this exciting, yet challenging, instrument is expected to commence this calendar year.

\acknowledgments
The LATR team and the SO collaboration would like to thank the CCAT-prime team for allowing us to borrow heavily from their telescope design. We thank the Vertex group who have been working closely with us on the telescope development.  Meyer Tool\footnote{Meyer Tool and Mfg., 4601 SW Hwy, Oak Lawn, IL 60453, USA, https://www.mtm-inc.com/}, Dynavac\footnote{Dynavac, 10 Industrial Park Rd \# 2, Hingham, MA 02043, USA, https://www.dynavac.com/}, PVeng, and Fermilab have provided a number of helpful suggestions regarding the cryostat design and fabrication process. Bluefors, Cryomech,  Oxford Instruments\footnote{Oxford Instruments, Tubney Woods, Abingdon, Oxfordshire, OX13 5QX, United Kingdom, https://www.oxford-instruments.com/}, and High Precision Devices (HPD) generously provided detailed cooling capacity information on their refrigeration units.  We also appreciate their help with the heat pipe and heat switch development process.  

This work was supported in part by a grant from the Simons Foundation (Award \#457687, B.K.).  RT acknowledges West Chester University of Pennsylvania for their support.



\def\ref@jnl#1{{\jnl@style#1}}

\def\aj{\ref@jnl{AJ}}                   
\def\actaa{\ref@jnl{Acta Astron.}}      
\def\araa{\ref@jnl{ARA\&A}}             
\def\apj{\ref@jnl{ApJ}}                 
\def\apjl{\ref@jnl{ApJ}}                
\def\apjs{\ref@jnl{ApJS}}               
\def\ao{\ref@jnl{Appl.~Opt.}}           
\def\apss{\ref@jnl{Ap\&SS}}             
\def\aap{\ref@jnl{A\&A}}                
\def\aapr{\ref@jnl{A\&A~Rev.}}          
\def\aaps{\ref@jnl{A\&AS}}              
\def\azh{\ref@jnl{AZh}}                 
\def\baas{\ref@jnl{BAAS}}               
\def\bac{\ref@jnl{Bull. astr. Inst. Czechosl.}}
\def\caa{\ref@jnl{Chinese Astron. Astrophys.}}
\def\cjaa{\ref@jnl{Chinese J. Astron. Astrophys.}}
\def\icarus{\ref@jnl{Icarus}}           
\def\jcap{\ref@jnl{J. Cosmology Astropart. Phys.}}
\def\jrasc{\ref@jnl{JRASC}}             
\def\memras{\ref@jnl{MmRAS}}            
\def\mnras{\ref@jnl{MNRAS}}             
\def\na{\ref@jnl{New A}}                
\def\nar{\ref@jnl{New A Rev.}}          
\def\pra{\ref@jnl{Phys.~Rev.~A}}        
\def\prb{\ref@jnl{Phys.~Rev.~B}}        
\def\prc{\ref@jnl{Phys.~Rev.~C}}        
\def\prd{\ref@jnl{Phys.~Rev.~D}}        
\def\pre{\ref@jnl{Phys.~Rev.~E}}        
\def\prl{\ref@jnl{Phys.~Rev.~Lett.}}    
\def\pasa{\ref@jnl{PASA}}               
\def\pasp{\ref@jnl{PASP}}               
\def\pasj{\ref@jnl{PASJ}}               
\def\rmxaa{\ref@jnl{Rev. Mexicana Astron. Astrofis.}}%
\def\qjras{\ref@jnl{QJRAS}}             
\def\skytel{\ref@jnl{S\&T}}             
\def\solphys{\ref@jnl{Sol.~Phys.}}      
\def\sovast{\ref@jnl{Soviet~Ast.}}      
\def\ssr{\ref@jnl{Space~Sci.~Rev.}}     
\def\zap{\ref@jnl{ZAp}}                 
\def\nat{\ref@jnl{Nature}}              
\def\iaucirc{\ref@jnl{IAU~Circ.}}       
\def\aplett{\ref@jnl{Astrophys.~Lett.}} 
\def\apspr{\ref@jnl{Astrophys.~Space~Phys.~Res.}}
\def\bain{\ref@jnl{Bull.~Astron.~Inst.~Netherlands}} 
\def\fcp{\ref@jnl{Fund.~Cosmic~Phys.}}  
\def\gca{\ref@jnl{Geochim.~Cosmochim.~Acta}}   
\def\grl{\ref@jnl{Geophys.~Res.~Lett.}} 
\def\jcp{\ref@jnl{J.~Chem.~Phys.}}      
\def\jgr{\ref@jnl{J.~Geophys.~Res.}}    
\def\jqsrt{\ref@jnl{J.~Quant.~Spec.~Radiat.~Transf.}}
\def\memsai{\ref@jnl{Mem.~Soc.~Astron.~Italiana}}
\def\nphysa{\ref@jnl{Nucl.~Phys.~A}}   
\def\physrep{\ref@jnl{Phys.~Rep.}}   
\def\physscr{\ref@jnl{Phys.~Scr}}   
\def\planss{\ref@jnl{Planet.~Space~Sci.}}   
\def\procspie{\ref@jnl{Proc.~SPIE}}   

\let\astap=\aap
\let\apjlett=\apjl
\let\apjsupp=\apjs
\let\applopt=\ao

\bibliography{ref} 
\bibliographystyle{spiebib} 

\end{document}